\documentclass[a4paper,11pt]{article}
\pdfoutput=1 

\usepackage{jheppub} 

\usepackage[T1]{fontenc} 

\usepackage{hyperref,amssymb,amsmath,mathrsfs,bm,graphicx}
\usepackage[dvipsnames]{xcolor}

\usepackage[dvipsnames]{xcolor}

\def\be{\begin{equation}}
\def\ee{\end{equation}}
\def\bea{\begin{eqnarray}}
\def\eea{\end{eqnarray}}

\def\yzero{\smash{\hbox{$y\kern-4pt\raise1pt\hbox{${}^\circ$}$}}}

\def\beq{\begin{equation}}
\def\eeq{\end{equation}}
\def\beqa{\begin{eqnarray}}
\def\eeqa{\end{eqnarray}}

\def\-{\hphantom{-}}

\def\s2{\frac{1}{\sqrt2}}

\def\beq{\begin{equation}}
\def\eeq{\end{equation}}
\def\beqa{\begin{eqnarray}}
\def\eeqa{\end{eqnarray}}

\def\IF{\relax{\rm I\kern-.18em F}}
\def\II{\relax{\rm I\kern-.18em I}}
\def\IP{\relax{\rm I\kern-.18em P}}
\def\IC{\relax\hbox{\kern.25em$\inbar\kern-.3em{\rm C}$}}
\def\IR{\relax{\rm I\kern-.18em R}}

\def\Dsl{\,\raise.15ex\hbox{/}\mkern-13.5mu D} 
\def\IZ{Z\kern-.4em  Z}

\title{\boldmath Worldsheet description of a \textit{massive} type IIA superstring in 10D}

\author[a]{M.P. Garcia del Moral}
\author[b]{P. Le\'on,\note{All authors have contributed equally to this work, and the names are listed in alphabetical order.}}
\author[c]{A. Restuccia}

\affiliation[a]{Área de Física, Departamento de Química, Centro Científico Tecnológico, \\ Universidad de la Rioja,La Rioja 26006, Spain}
\affiliation[b]{Perimeter Institute for Theoretical Physics, \\ Waterloo, ON N2L 2Y5, Canada.}
\affiliation[c]{Departamento de F\'isica, Universidad de Antofagasta,\\Aptdo 02800,  Chile.}

\emailAdd{m-pilar.garciam@unirioja.es}
\emailAdd{pleon@perimeterinstitute.ca}
\emailAdd{alvaro.restuccia@uantof.cl}

\abstract{We construct, following \cite{mpgm14,mpgm17}, a massive M2-brane (supermembrane) as the limit of a genus two M2-brane that becomes a twice punctured Riemann surface with particular boundary conditions on the fields defined on the punctures. The target space is $M_9\times LCD$, where $LCD$ is a genus one light cone diagram. It contains mass terms and a topological term associated with the non-triviality of the target surface that, at low energies, can be associated with the presence of a cosmological constant. 
We show that the supergravity background of the M2-brane considered in this formulation requires the presence of M9-branes acting as sources. They correspond to the 11D uplift of the characteristic D8's of Romans supergravity. To this end, we explicitly show that some of the background singularities of the massive M2-brane can be reproduced by the M9-branes found by \cite{Bergshoeff}. This establishes a relation between the Romans mass and the moduli of the massive M2-brane.
 
When dimensionally reduced, we obtain a worldsheet Hamiltonian of a N=2 type IIA closed superstring in 10D. We denote it \textit{massive} string. The corresponding \textit{massive} string inherits a non-vanishing constant term from the topological massive M2-brane that shifts the Hamiltonian. The non-vanishing parameter is related to the non-trivial structure of the massive M2-brane background and it can be related to the Romans mass term. It also contains a modified tension due to the non-trivial dependence on the moduli and on the punctures associated with the target torus.}

\begin{document} 
\maketitle
\flushbottom

\section{Introduction}

The superstring theory in 10D and its nonperturbative description, M-theory in 11D are unification theories of all the fundamental interactions in a single framework.  The low energy limit of M-theory on a flat space corresponds to the maximal supergravity in eleven dimensions. From the 11D supergravity formulation, it is possible to obtain, through a Kaluza Klein reduction, the maximal IIA supergravity in 10D, which is the low energy limit of the $N=2$ type IIA superstring. A Scherk-Schwarz's reduction of the 11D supergravity leads to a gauged deformation of the type IIA supergravity \cite{Howe}. On the other hand, in 10D, the type IIB sector is obtained through a T-duality transformation. In \cite{Romans}, it was found that there exists a massive type IIA supergravity in 10D, known as Romans supergravity, whose origin in M-theory was not clear. Brane solutions associated with Type IIA massive supergravity were found in \cite{Janssen}.  In \cite{mpgm13} a M-theory origin proposal was made. Its uplift to eleven dimensions supergravity had been previously established in \cite{Bergshoeff6}. There, the authors found that the corresponding 11D action couples to M9-branes, which are the uplift of the 10D $D8$-branes characteristics of Romans supergravity. The 11D massive supergravity exhibits a cosmological term associated with the Romans supergravity mass parameter that makes the theory non-covariant. In order to include consistently the cosmological term in the 11D action a killing isometry in the space-time is required \cite{Bergshoeff6}.

Besides, in \cite{Aharony2}, the authors showed that for particular backgrounds, the massive type IIA string cannot be strongly coupled, which they took as an indication that this theory could not have an origin in M-theory. However their argument was done in a weakly curved region of space-time, something that the model that we present evades. 

In \cite{Hull8}, it was conjectured that the M-theory origin of type IIA massive Romans supergravity could be obtained by taking the 11D uplift of the M-theory formulated on a T-dual torus bundle with parabolic monodromy. A matrix theory approach of Hull’s conjecture was proposed in \cite{Lowe}. A realization of this idea in terms of supermembranes was proposed in \cite{mpgm13},  where the non-trivial uplift of the M2-brane with central charge formulated on a punctured torus bundle  and parabolic monodromy was obtained. We will use interchangeably throughout the work the name supermembrane or M2-brane. Following this idea we constructed, in \cite{mpgm14}, a rigorous formulation of the M2-brane on $M_9\times LCD$, where $LCD$ is a Light Cone Diagram, a two dimensional flat strip with identifications and with prescribed segments whose curvature becomes infinite at some points. The Mandelstam map, see  \cite{Giddings2,Mandelstam,Mandelstam3,Restuccia5}, realizes the conformal equivalence between the punctured Riemann surface and the $LCD$. The real part of the map corresponds to the Green function of the Laplace operator on the surface. It is the conformal time on the surface, a single-valued function which maps the punctures to infinity on the $LCD$. At supergravity level, the punctures in Riemann surfaces lead to delta function singularities in the equations of motion.  These singularities can be associated to the existence of Dp-brane sources \cite{Hull10}, where $p$ depends on the dimension of the delta function. In the case of the massive Roman supergravity in ten dimensions and its  uplift to eleven dimensions, as we already mentioned, they have been related to the coupling with D8-branes and M9-branes, respectively \cite{Bergshoeff7,Sato}.  In \cite{Bah,Bobev,Lozano,Dibitetto} are considered examples of massive type IIA supergravity on punctured Riemann surfaces generated by Dp-branes or M2-branes stacks.

The formulation of the massive supermembrane in \cite{mpgm14,mpgm17}  exhibits several features associated with the non-trivial topology of the target space that are not present in a standard compactification. The Hamiltonian contains a non-trivial cosmological term and  a bosonic potential with new non-vanishing quadratic mass terms associated with all dynamical fields. These quadratic terms, together with the rest of the contributions of the supersymmetric potential, ensure that the potential does not contain valleys in any direction. Thus, provided a proper matrix model regularization for the Hamiltonian, the regularized theory satisfies the sufficient condition for discreteness \cite{mpgm11}  and the supersymmetric spectrum must be discrete \cite{mpgm14}. Therefore, the M2-brane on a punctured Riemann surface we propose is one of the few known cases where the Hamiltonian of the theory has this relevant property.  In comparison to a standard $S^1$ compactification of the theory, the Hamiltonian of the M2-brane on $M_9\times LCD$ target space is subject to more constraints, associated with the area preserved diffeomorphisms that leave the punctures fixed.

In \cite{Cattabriga} it was shown that there exists a correspondence between the mapping class group of the twice punctured torus and the (1,1)-knots. This result allowed us to classify the monodromies over the twice-punctured torus (and thus the monodromies of the M2-brane on $M_9\times LCD$) in terms of (1,1)-knots.  In order to interpret the supermembrane on $M_9\times LCD$ as the non-trivial uplift to ten non-compact dimensions of the M2-brane on a torus bundle with $C_-$ fluxes, the monodromies must be restricted to the subgroup of the mapping class group of twice-punctured torus that preserves the decompactified direction. 
To achieve this result, the unique monodromies allowed are the parabolic ones, which is consistent with the results of \cite{mpgm12} and the conjecture made by C. Hull in \cite{Hull8}.
 
In this work, we study the reduction of the supermembrane on $M_9\times LCD$ in order to obtain the worldsheet Hamiltonian of a \textit{massive} type IIA superstring in ten non-compact dimensions. Now, the dimensional reduction must be performed in such a way that the topological terms resulting from the compactification on the $LCD$ are preserved. Hence, we will not follow the standard double dimensional reduction (see for example \cite{Duff}). Instead, we will use the symmetries of the theory to fix a gauge (similar to the KK reduction) to reduce the target-space. Then, the reduction of the worldvolume will follow from the properties of the Riemann surfaces with punctures. It is also necessary to reduce the number of constraints in the theory. In this sense, the properties of the Mandelstam map will play a fundamental role. We organize this work in the following way:  Section 2 summarizes the key findings of our previous work done in \cite{mpgm14} by presenting the Hamiltonian of the M2-brane on $M_9\times LCD$. In section 3, we  discuss the dimensional reduction of the theory and its constraints and we obtain the action of the massive type IIA string theory in ten non-compact dimensions. Finally, in section 4, we analyze the results obtained of the work.
\section{The supermembrane action in the light cone gauge}

The supermembrane was originally introduced in \cite{Bergshoeff}. Its formulation in the Light Cone Gauge (LCG) on a Minkowski target space was obtained in \cite{dwhn}.
In this section we will briefly review some of those results \cite{dwhn} and we will present the supermembrane action  in the light cone gauge on $M_9 \times T^2$. The action of the supermembrane in a Minkowski space-time is given by 

\begin{eqnarray}
S &=& -T_{M2} \int_{R\times \Sigma} d\xi^3 \bigg[\sqrt{-g} +\varepsilon^{uvw} \bar{\tilde{\Psi}} \Gamma_{\mu \nu} \partial_{w}\tilde{\Psi}  \nonumber \\ &\times&\bigg(\frac{1}{2}\partial_u \tilde{X}^\mu (\partial_v \tilde{X}^\nu + \bar{\tilde{\Psi}}\Gamma^\nu \partial_v \tilde{\Psi}) + \frac{1}{6} \bar{\tilde{\Psi}}\Gamma^\mu \partial_u \tilde{\Psi} \bar{\tilde{\Psi}}\Gamma^\nu\partial_v\tilde{\Psi}\bigg)\bigg], \nonumber \\ &&
\end{eqnarray}
where $T_{M2}$ is the M2-brane tension, $\Gamma^\mu$ are the gamma matrix in eleven dimensions, $\tilde{X}^\mu$ $(\mu,\nu = 0,..,10)$ are the embedding maps of the supermembrane, $\theta$ is a 32 component Majorana spinor and $\Sigma$ is a compact Riemann surface. All the fields are functions of the worldvolume coordinates $\xi^u$ $(u,v,w = 0,1,2)$ and $g_{uv}$ are the components of the worldvolume induced metric, this is

\begin{eqnarray}
g_{uv} = (\partial_u \tilde{X}^\mu + \bar{\tilde{\Psi}}\Gamma^\mu \partial_u \tilde{\Psi})(\partial_v \tilde{X}^\nu + \bar{\tilde{\Psi}}\Gamma^\nu \partial_v \tilde{\Psi}) \eta_{\mu \nu}.
\end{eqnarray}

Now we can use the light cone coordinates $\tilde{X}^\mu = (X^+,X^-,\tilde{X}^{M})$ with $M,N = 1,..,9$

\begin{eqnarray}
X^{\pm} = \frac{1}{\sqrt{2}}(\tilde{X}^{10} \pm \tilde{X}^0), \quad \Gamma^{\pm} = \frac{1}{\sqrt{w}}(\Gamma^{10} \pm \Gamma^0),
\end{eqnarray}
and, by decomposing $\xi^u=(t,\sigma^r)$ with $r=1,2$, one can fix the LCG as follows,

\begin{eqnarray}
X^+ = t, \quad \Gamma^+ \tilde{\Psi} = 0. 
\end{eqnarray}
Thus, following the notation introduced in \cite{dwhn} we can write the Lagrangian density can be written as \footnote{We are using $\varepsilon^{0rs} = -\epsilon^{rs}$}

\begin{eqnarray}
\mathcal{L} &=& -T_{M2} (\sqrt{\bar{g}\Delta} + \epsilon^{rs}\partial_r \tilde{X}^{M} \bar{\tilde{\Psi}}\Gamma^-\Gamma_M\partial_s \tilde{\Psi}),
\end{eqnarray}
where 

\begin{eqnarray*}
\bar{g}_{rs} &=& \partial_r \tilde{X}^{M} \partial_s \tilde{X}_M, \quad 
u_r = g_{0r} = \partial_r \tilde{X}^- + \partial_t \tilde{X}^{M}\partial_r \tilde{X}_M + \tilde{\Psi}\Gamma^-\partial_r \tilde{\Psi}, \\
g_{00} &=& 2\partial_0X^-+\partial_t \tilde{X}^{M}\partial_t\tilde{X}_M + 2\bar{\tilde{\Psi}}\Gamma^-\partial_0\tilde{\Psi},
\end{eqnarray*}
and $\bar{g} = det(\bar{g}_{rs})$, $\Delta = -g_{00} + u_r \bar{g}^{rs}u_s$. Then, the conjugate momenta can be written as

\begin{align*}
\small
& \tilde{P}_- = T_{M2} \sqrt{\frac{\bar{g}}{\Delta}}, \quad \tilde{P}^M = \tilde{P}_- (\partial_0\tilde{X}^{M}-u_rg^{rs}\partial_s\tilde{X}^{M}), \quad \tilde{S} = - \tilde{P}_-\Gamma^-\tilde{\Psi}.
\end{align*}
Thus, the Hamiltonian density is given by

\begin{eqnarray}
\mathcal{H} = \frac{\tilde{\textbf{P}}^2+T^2_{M2}\bar{g}}{2\tilde{P}_-} - T_{M2}\epsilon^{rs}\partial_r\tilde{X}^{M} \bar{\tilde{\Psi}} \Gamma^-\Gamma_M\partial_s \tilde{\Psi},
\end{eqnarray}
subject to the following constraints 
\begin{align}\label{fc0}
 &  \tilde{\textbf{P}} \partial_r \tilde{\textbf{X}} + \tilde{P}_-\partial_r \tilde{X}^- + \bar{\tilde{S}}\Gamma^-\tilde{\Psi} = 0, \quad \tilde{S} + \tilde{P}_-\Gamma^-\tilde{\Psi} = 0.
\end{align}
 Now, we can use the area preserving diffeomorphims to set the gauge $\tilde{P}_- = P_-^0\sqrt{W}$, where $\sqrt{\tilde{W}}$ is a scalar density satisfying 

\begin{eqnarray}
\int_\Sigma \sqrt{\tilde{W}} = 1.
\end{eqnarray}
This allows to introduce the Lie bracket

\begin{eqnarray}
\{\bullet, \bullet \} = \frac{\epsilon^{rs}}{\sqrt{\tilde{W}}}\partial_r \bullet \partial_s \bullet.
\end{eqnarray}

The supermembrane Lagrangian density can be written in a way that is explicitly invariant under area preserving diffeomorphims (see \cite{dwhn}). This requires the introduction of a gauge field $\omega$ related to time-dependent reparametrizations of the worldvolume. This will be
\begin{align}\label{SL}
\frac{\mathcal{L}}{P_0^+ \sqrt{\tilde{W}}} = \frac{1}{2}(D_0\tilde{X}^{M})^2+\bar{\tilde{\Psi}}\Gamma^- D_0\tilde{\Psi} - \frac{T_{M2}^2}{4P_0^+}\{\tilde{X}^{M}, \tilde{X}^N\}^2  + \frac{T_{M2}}{P_0^+}\bar{\tilde{\Psi}}\Gamma^-\Gamma^a \{\tilde{X}^{M},\tilde{\Psi}\} +D_0 \tilde{X}^-,    
\end{align}
where

\begin{eqnarray}
   D_0 \bullet = \partial_t \bullet - \{\omega, \bullet\},
\end{eqnarray}
Furthermore, we can now solve (\ref{fc0}) for $\tilde{X}^-$, this is 

\begin{eqnarray}
\partial_r \tilde{X}^- = -\frac{1}{P_-^0\sqrt{\tilde{W}}} (\tilde{\textbf{P}} \partial_r \tilde{\textbf{X}}+\bar{\tilde{S}}\Gamma^-\partial_r\tilde{\Psi}).
\end{eqnarray}
In order to ensure the existence of a single-valued solution for $\tilde{X}^-$, one must impose

\begin{eqnarray}
 \phi &=& d(d\tilde{X}^-) = d\left[\frac{1}{\sqrt{\tilde{W}}} (\tilde{\textbf{P}} d \tilde{\textbf{X}}+\bar{\tilde{S}}\Gamma^-d\tilde{\Psi})\right] = 0 \\
 \varphi_k &=& \int_{\mathcal{C}_k} d\tilde{X}^- = \int_{\mathcal{C}_k}\frac{1}{\sqrt{\tilde{W}}} (\tilde{\textbf{P}} d \tilde{\textbf{X}}+\bar{\tilde{S}}\Gamma^-d\tilde{\Psi}) =0,
\end{eqnarray}
where $\mathcal{C}_k$ ($k = 1,..,2g$ for $g>1$) are the homology basis of one-cycles over $\Sigma$. 
They correspond to the local and global first class constraints associated with the residual symmetry of Area Preserving Diffeomorphisms (APD).

It is possible now, to write the Hamiltonian of the theory as,

\begin{align}\label{g2H}
H = \frac{1}{2P_-^0} \int_\Sigma d^2\sigma \sqrt{\tilde{W}} \bigg[\left(\frac{\tilde{\textbf{P}}}{\sqrt{\tilde{W}}}\right)^2 + \frac{T^2_{M2}}{2}\{\tilde{X}^{M},\tilde{X}^N\}^2 - 2T_{M2} P_-^0 \bar{\tilde{\Psi}}\Gamma^-\Gamma_M \{\tilde{X}^{M},\tilde{\Psi}\}\bigg].    
\end{align}

Then, one can compactify the M2-brane Hamiltonian on $M_9 \times T^2$ and take as a base manifold a regular genus-two Riemann surface $\Sigma_2$. Thus, due to the compact dimensions, the embedding maps can be decomposed as $\tilde{X}^{M} = (\tilde{X}^m,\tilde{X}^r)$, with $m=1,...,7$ labelling the non-compact dimensions and $r=1,2$ the compact ones associated with the 2-torus. The $\tilde{X}^m$ maps $\Sigma_2$ to the transverse subspace of $M_9$ while $\tilde{X}^r$ maps $\Sigma_2$ to the target $T^2$.

 Hence, the Hamiltonian of the supermembrane can be written as

\begin{align}
   &H = \frac{1}{2P_-^0} \int_{\Sigma_2} d^2\sigma \sqrt{\tilde{W}} \bigg[\left(\frac{\tilde{P}_m}{\sqrt{\tilde{W}}}\right)^2 + \left(\frac{\tilde{P}_r}{\sqrt{\tilde{W}}}\right)^2 + \frac{T^2_{M2}}{2}\{\tilde{X}^m,\tilde{X}^n\}^2 + T^2_{M2}\{\tilde{X}^m,\tilde{X}^r\}^2 \nonumber \\ &    + \frac{T^2_{M2}}{2}\{\tilde{X}^r,\tilde{X}^s\}^2 -2T_{M2} P_-^0 \bar{\tilde{\Psi}}\Gamma^-\Gamma_m \{\tilde{X}^m,\tilde{\Psi}\} - 2T_{M2} P_-^0 \bar{\tilde{\Psi}}\Gamma^-\Gamma_r \{\tilde{X}^r,\tilde{\Psi}\}\bigg], 
\end{align}
 subject now to the following  five APD constraints

\begin{eqnarray}
 \phi &=& d\left[\frac{1}{\sqrt{\tilde{W}}} (\tilde{P}_m d\tilde{X}^m +\tilde{P}_r d\tilde{X}^r + \bar{\tilde{S}}\Gamma^-d\tilde{\Psi})\right] = 0, \\ 
 \varphi_k &=& \int_{\mathcal{C}_k}\frac{1}{\sqrt{\tilde{W}}} (\tilde{P}_m d\tilde{X}^m +\tilde{P}_r d\tilde{X}^r +\bar{\tilde{S}}\Gamma^-d\tilde{\Psi}) =0, 
\end{eqnarray}
where $k=1,...,4$. We will use these expressions in the subsequent sections of the paper. 
\section{The massive supermembrane on $M_9\times LCD$}

In this section, we will give a brief review of the results about the formulation of the massive supermembrane found in \cite{mpgm17}. The specific limit of a supermembrane on $M_9 \times T^2$ over a compact Riemann surface of genus two $\Sigma_2$ yields this supermembrane formulation. Specifically, when one of the handles of $\Sigma_2$ shrinks to a string. In this limit, we end up with a twice punctured torus with a string attached to it. Under the right conditions for the supermembrane maps, the string does not carry M2-brane energy and, therefore, is one of the string-like configurations discussed in \cite{Nicolai,dwln}. This particular construction for the massive supermembrane has two main advantages. The first one is that it enables us to use all of the results from \cite{mpgm10}. The second advantage is that it allows for a rigorous treatment of all of the theory's surface terms as well as the supersymmetric algebra (see \cite{mpgm17} for more details). Furthermore, we can keep the interpretation of this formulation as the decompactification limit of the supermembrane on a torus bundle with parabolic monodromy and therefore an explicit realization of Hull's conjecture \cite{Hull8}. It imposes boundary conditions on the fields valued at the punctures.

\subsection{The twice punctured torus}

\begin{figure*}
    \centering
    \includegraphics[scale=0.4]{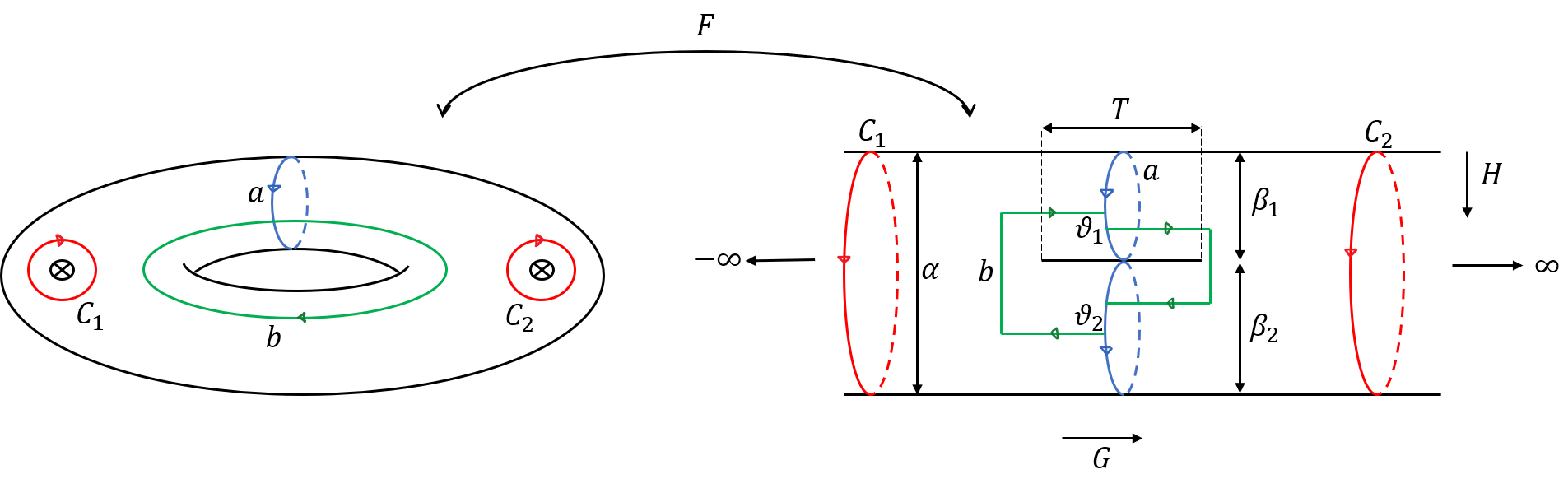}
    \caption{The torus with two punctures and the one loop interaction string diagram with one incoming/outgoing string. The Mandelstam map send the punctures over the torus to $\pm\infty$ in the $LCD$.}
    \label{fig:equi}
\end{figure*}

Here we will review some useful results about the relation between the Light Cone Diagram ($LCD$) and the torus with two punctures $\Sigma_{1,2}$ (see figure (\ref{fig:equi}))  needed to describe the massive supermembrane formulation. The Light Cone diagram is a two dimensional flat strip with identifications and with
prescribed segments whose curvature becomes infinite at some points. These results are the base of the massive supermembrane formulation \cite{mpgm14} and they will be useful in the next sections. The relation between these two surfaces is given by the Mandelstam map (see \cite{Mandelstam,Giddings2})

\begin{eqnarray}\label{Mm}
    F(z)&=&\alpha \ln\left[\frac{\Theta_1(z-Z_1\vert \tau)}{\Theta_1(z-Z_2\vert \tau)}\right]-2\pi i \alpha\frac{Im (Z_1-Z_2)}{Im \tau}(z-z_0), \nonumber \\ &&
\end{eqnarray}
where $\Theta_1(z,\tau)$ are the Jacobi functions and $Z_r$ with $r=1,2$ are the positions of the punctures in a complex coordinates over the torus. The set of parameters necessary to characterize the torus with two punctures are the Teichm\"uller parameter, $\tau$, and the positions of the Punctures, $Z_r$. On the twice punctured torus the coordinate system $z$ is defined in terms of the holomorphic one-form $dz$ satisfying

\begin{eqnarray}
dz = d\hat{X}^1 + \tau d\hat{X}^2,
\end{eqnarray}
where $d\hat{X}^r$ is a set of real normalized forms over the regular torus, this is

\begin{eqnarray}
\int_{\mathcal{C}_k} d\hat{X}^r = \delta_k^r.
\end{eqnarray}

On the other hand, the set of parameters that describe the $LCD$ are the external weights $\alpha$, the internal weights $\beta_r$, the internal length $T$, and the twist angles $\theta_r$. This parameter can be written in terms of the Mandelstan map as following,

\begin{eqnarray}
    &&  \int_{C_r} dF = (-1)^{r+1}2\pi i \alpha, \quad \int_a dF = 2\pi i \beta_1, \quad \int_b dF = \frac{i}{2\pi}(\beta_1 \theta_1-\beta_2 \theta_2), 
\end{eqnarray}

\begin{eqnarray} \label{itt}
    T \equiv \int_{P_1}^{P_2}dG.
\end{eqnarray}
Then in order to complete the equivalence between the two surfaces, (see figure 1.),  the following relation between both sets of parameter is required 

\begin{equation}
2\pi i (Z_1-Z_2)= (\theta_1+\theta_2)\beta_1-\alpha\theta_2-2\pi i\beta_1 \tau.
\end{equation}  
It is useful to decompose the Mandelstam map in terms of its real and imaginary parts, that is $F= G+iH$. The function $G$ is single valued, but $dG$ is harmonic, since it has poles at the punctures. The function $H$ is multivalued and $dH$ is harmonic. The behavior of each function near the punctures is given by
\begin{eqnarray}
G  &\sim&  (-1)^{r+1}\alpha\ln\vert z-Z_r\vert, \\
H &\sim& (-1)^{r+1}\alpha \varphi, \quad \mbox{with} \quad \varphi\in [0,2\pi] \ \ (r=1,2). \label{HB}
\end{eqnarray}
On the other hand, near the zeros of $dF$, denoted as $P_a$, the functions $G$ and $H$ can be written as
\begin{eqnarray}
G(z)-G(P_a) &\sim &  \frac{1}{2}Re(D(P_a)(z-P_a)^2),\label{Gz}  \\
H(z)-H(P_a)  & \sim &  \frac{1}{2}Im(D(P_a)(z-P_a)^2), \label{Hz}
\end{eqnarray}
where
\begin{align}
    D(P_a) = \sum_{r=1}^{2} (-1)^{r+1} \bigg[ \frac{\partial^2_z \Theta_1(P_a-z_r,\tau)}{\Theta_1(P_a-z_r,\tau)}-\left(\frac{\partial_z \Theta_1(P_a-z_r,\tau)}{\Theta_1(P_a-z_r,\tau)}\right)^2 \bigg].   
\end{align}
Finally, we recall some  properties of the functions $K$ and $H$ that will be useful in the next section,

\begin{eqnarray}
    G(z+1)-G(z) &=& G(z+\tau)-G(z)=0  \label{Ks}\\
    H(z+1)-H(z) &=& 2\pi \alpha \frac{Im(Z_2-Z_1)}{Im(\tau)}, \label{Hs1} \\
    H(z+\tau)-H(z) & = & \frac{2\pi \alpha Im((Z_2-Z_1)\bar{\tau})}{Im(\tau)}. \label{Hs2}
\end{eqnarray}

\subsection{Massive supermembrane}

\begin{figure*}
    \centering
    \includegraphics[scale=0.4]{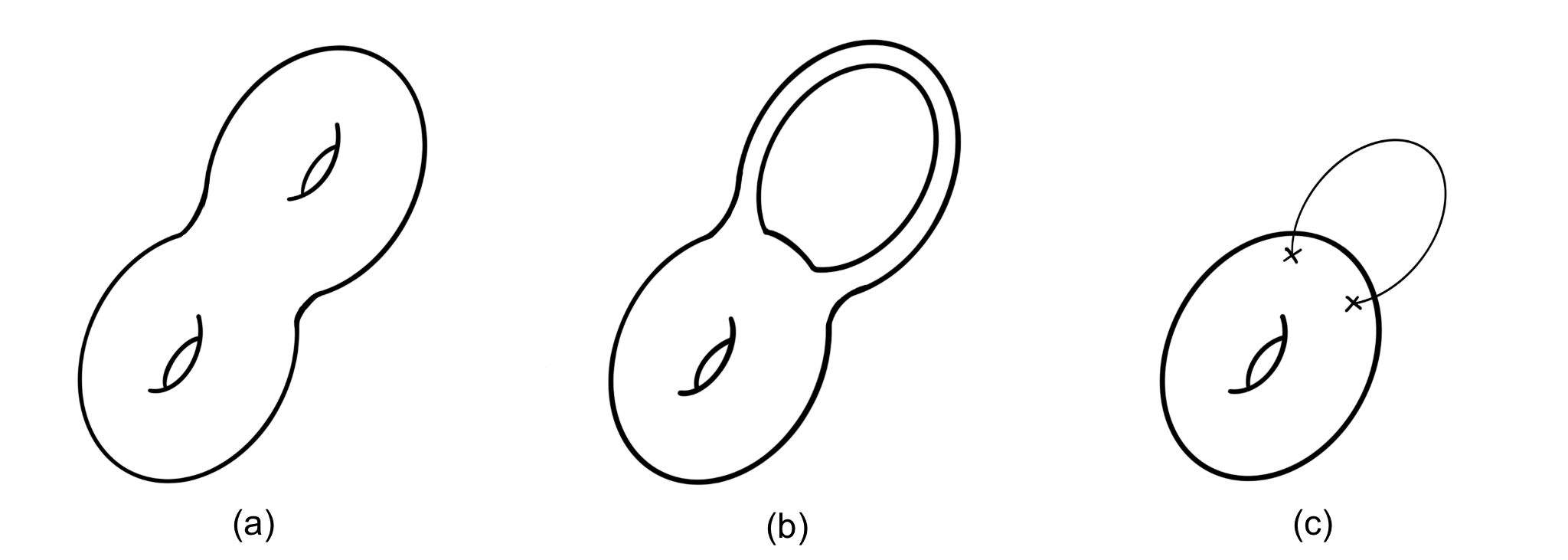}
    \caption{(a) The genus two regular Riemann surface $\Sigma_2$. (b) A deformation of $\Sigma_2$. (c) The surface $\tilde{\Sigma}_{1,2}$ obtained by taking one of the radii of $\Sigma_2$ tending to zero. This corresponds to a singular $T^2$  with a string attached to it.}
    \label{fig:TD}
\end{figure*}

Following the procedure developed in \cite{mpgm17} we will begin with the formulation of the supermembrane on compact genus-two Riemann surface $\Sigma_2$ as the base manifolds and $M_9\times T^2$ as the target space. Then we will deform $\Sigma_2$ as described in the figure (\ref{fig:TD}). That is, we will take the limit in which one of the radii of the handles of the genus two surface tends to zero. Thus, we can expand the maps $X^m$, $X^r$ and $\Psi$, in a Fourier series and keep only the order zero of the variable associated with the small radius. This implies that, in this limit, all the supermembranes fields will depend only on the coordinate along the handle (see figure (\ref{fig:TD})-(b)). In this way, we get a string-like configuration like the ones described in \cite{Nicolai} and the final surface, that will denote as $\tilde{\Sigma}_{1,2}$, is essentially a twice punctured torus $\Sigma_{1,2}$ with a string attached to the punctures (see figure (\ref{fig:TD})-(c)). 

Now, we can describe the dependence of the M2-brane fields in two regions. The first one is the definition of the maps on $\Sigma_{1,2}$, where we can follow the same definition used in \cite{mpgm14}. The second region is the string attached to $\Sigma_{1,2}$, that we shall denote as $\gamma_2$. Then,  given a coordinate system, $z$ (given in the previous section), over $\Sigma_{1,2}$ and defining as $u$ the coordinate associated with $\gamma_2$, we can write

\begin{eqnarray}
    \tilde{X}^m = \left\lbrace 
    \begin{array}{ll}
        X^m(t,z,\bar{z}) & \mbox{over} \quad \Sigma_{1,2}  \\
        Y^m(t,u) &  \mbox{over} \quad \gamma_2
    \end{array}
    \right. ,  \quad
     \tilde{\Psi} = \left\lbrace 
    \begin{array}{ll}
        \Psi(t,z,\bar{z}) & \mbox{over} \quad \Sigma_{1,2}  \\
        \Theta(t,u) &  \mbox{over} \quad \gamma_2
    \end{array}
    \right. ,
\end{eqnarray}
and
\begin{eqnarray}
    \tilde{X}^r = \left\lbrace 
    \begin{array}{ll}
        X^K(t,z,\bar{z})\delta^r_1 +  X^H(t,z,\bar{z})\delta^r_2& \mbox{over} \quad \Sigma_{1,2}  \\
        Y^r(t,u) &  \mbox{over} \quad \gamma_2
    \end{array}
    \right. .
\end{eqnarray}
The maps $X^K$ and $X^H$ are defined as in \cite{mpgm14}, i.e, 

\begin{eqnarray}
    X^K = K+A^K, \quad X^H = m H + A^H,
\end{eqnarray}
where $m$ is an integer, that outside the punctures can be interpreted as a winding, and the 1-forms $dA^K$, $dA^H$ are exact over $\Sigma_{1,2}$.
Then, we deform the target $T^2$ into a $LCD$ surface. Thus, the metric that we define over the $LCD$ on the target is given by

\begin{eqnarray}
     ds^2 &=& \frac{l^2}{\cosh^4(\hat{G})}d\hat{G}^2+ dH^2 = dK^2 + \alpha^2 d\hat{H}^2 \label{mono0},
\end{eqnarray}
where $\hat{H} = H/\alpha, \quad \hat{G} = G/\alpha$ and $l$ is constant with length units. Now, since the constant $l$ is associated with the coordinate $G$, it is reasonable to assume that it should be related with the only characteristic horizontal length of the $LCD$, this is,

\begin{eqnarray}
    l = \frac{k T}{\alpha} = k \hat{T},
\end{eqnarray}
where $\hat{T}\equiv T/\alpha$ and $k$ is constant  with length units. The reason why we introduce the constant $k$ will be clear in the last section of this work in which $\alpha$ plays an important role. Notice, from (\ref{Mm}) and (\ref{itt}), that $l$ is independent of $\alpha$.

 Under all these considerations, and as discussed in \cite{Nicolai,dwln}, the string attached to $\Sigma_{1,2}$ does not change the supermembrane energy and therefore we can write
\begin{eqnarray}
H = \int_{\Sigma_2 \rightarrow \tilde{\Sigma}_{1,2}}\mathcal{H} = \int_{\Sigma_{1,2}}\mathcal{H}.
\end{eqnarray}
Moreover, since the string-like configuration that we are considering here has no M2-brane dynamics associated with it, without losing generality we can impose

\begin{eqnarray} \label{cts}
   && Y^m_s(u,t) = const, \quad Y^r_s(u,t) = const, \quad \Theta_s(u,t)=const, \nonumber \\
   &&   
\end{eqnarray}
which implies 

\begin{eqnarray} \label{mbc1}
    X^m\bigg|_{Z_1}^{Z_2} = \Psi \bigg|_{Z_1}^{Z_2} = 0.
\end{eqnarray}
On the other hand, since the $Y^r_s(u,t)$ are singled-valued functions and $dA^K$ and $dA^H$ are exact 1-forms over $\Sigma_{1,2}$, it is reasonable to consider 

\begin{eqnarray} \label{mbc2}
    A^K\bigg|_{Z_1}^{Z_2} = A^H \bigg|_{Z_1}^{Z_2} = 0.
\end{eqnarray}

At this point we can follow the same steps presented in \cite{mpgm14} to analyse the Hamiltonian over $\Sigma_{1,2}$. Specifically, we define the worldvolume metric, over $\Sigma_{1,2}$, as 

\begin{eqnarray}
     \sqrt{W} &=& \frac{1}{4\pi}\epsilon^{rs}\partial_r \hat{K} \partial_s \hat{H},
\end{eqnarray}
where $K \equiv \tanh{\hat{G}}$. Then we can fix the gauge 

\bea
\{K,A^K\}+ m\{H,A^H\}=0. \label{gc}
\eea 

\begin{figure*}
    \centering
    \includegraphics[scale=0.1]{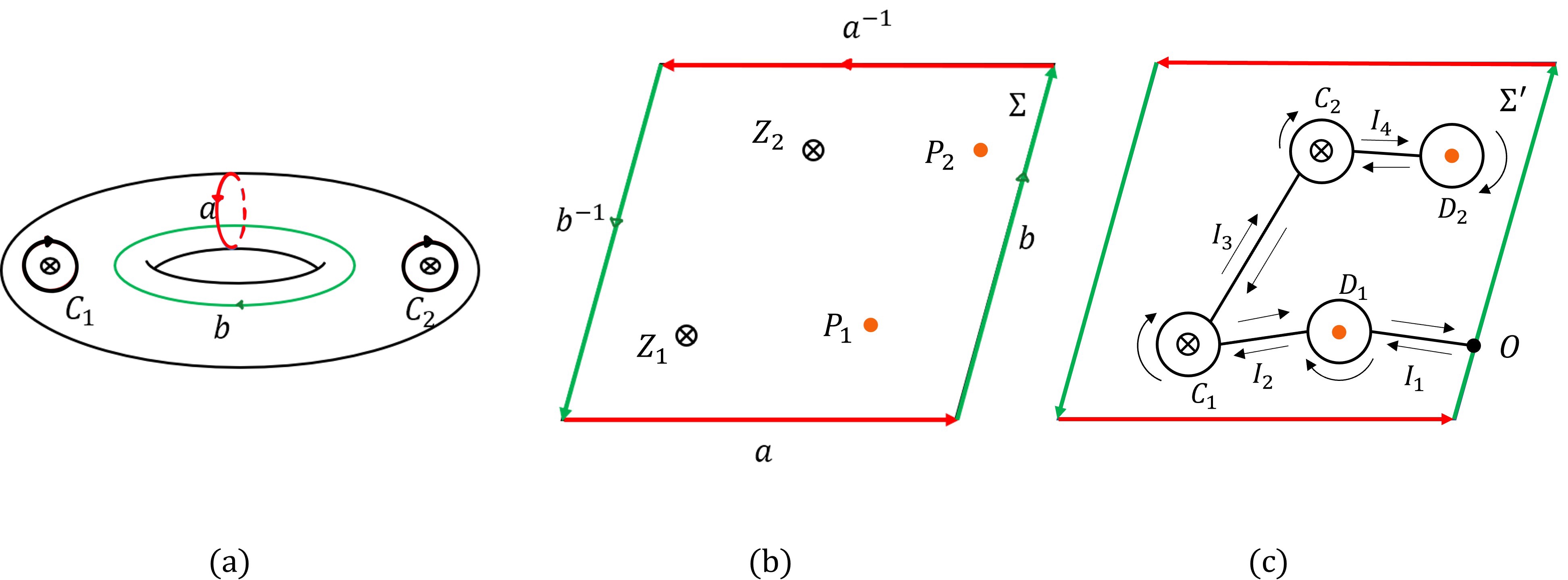}
    \caption{The region $\Sigma'$ obtained by cutting $\hat{\Sigma}_{1,2}$ through the curves $C_1$,$C_2$, $D_1,D_2$ and $I$. The path obtained by the union of the curves $C_1$,$I$,$C_2$,$D_1$$D_2$ and $I^{-1}$ is denoted by $c$}
    \label{fig:SigmaP}
\end{figure*}

We denote $\hat{\Sigma}_{1,2}$ the fundamental region in the complex plane of the punctured Riemann surface $\Sigma_{1,2}$. In order to deal with the singular behavior of the metric at the punctures and zeros we cut $\hat{\Sigma}_{1,2}$ through a closed curve that circumvents the two punctures, and the zeros with a radius $\epsilon$ and touch a point in the boundary of $\hat{\Sigma}_{1,2}$, see figure \ref{fig:SigmaP} (see \cite{farkas}). We denote as $C_r$ the curves around the punctures, $D_r$ the curves around the zeros, and $I_j$, with $j=1,..,4$, to the curves between them. Following the discussion presented in \cite{mpgm14}, it is clear that the curves $I_j$ can be chosen as curves $H=cte$.  We denote as $\Sigma'$ the resulting region after cutting $\hat{\Sigma}_{1,2}$.

Under all these considerations, the Hamiltonian of the theory can be written as (see \cite{mpgm14} for more details)

\begin{align}
    H &= \frac{(l  \alpha T_{M2} m )^2}{2P_0^+} +  \frac{1}{2P_0^+}\lim_{\epsilon\rightarrow 0} \int_{\Sigma'}d\sigma^2\sqrt{W}\left[\left(\frac{P_m}{\sqrt{W}}\right)^2 +\left(\frac{P_K}{\sqrt{W}}\right)^2 + \left(\frac{P_H}{\sqrt{W}}\right)^2 \right.\nonumber \\
    &+ T_{M2}^2\bigg(\frac{1}{2}\{X^m,X^n\}^2 + m^2\{X^m,H\}^2 + \{X^m,K\}^2 + 2\{X^m,K\}\{X^m,A^K\} \nonumber \\ &   + \{X^m,A^K\}^2 +  2m\{X^m,H\}\{X^m,A^H\} +\{X^m,A^H\}^2 + 2m\{H,A^K\}\{A^H,A^K\} \nonumber \\ & + m^2\{H,A^K\}^2 + 2\{A^H,K\}\{A^H,A^K\} + \{A^H,A^K\}^2  + \{K,A^H\}^2 +  \{K,A^K\}^2 \nonumber \\ & + \{H,A^H\}^2 \bigg)  -2P_0^+ T_{M2}  (\bar{\Psi}\Gamma^-\Gamma_m \{X^m,\Psi\}+ \bar{\Psi}\Gamma^-\Gamma_K \{A^K,\Psi\}+\bar{\Psi}\Gamma^-\Gamma_H \{A^H,\Psi\} \nonumber \\ & + \bar{\Psi}\Gamma^-\Gamma_K\{K,\Psi\} +  \bar{\Psi}\Gamma^-\Gamma_H\{H,\Psi\}) \Bigg]\label{MM}.
\end{align}

\begin{figure*}
    \centering
    \includegraphics[scale=0.4]{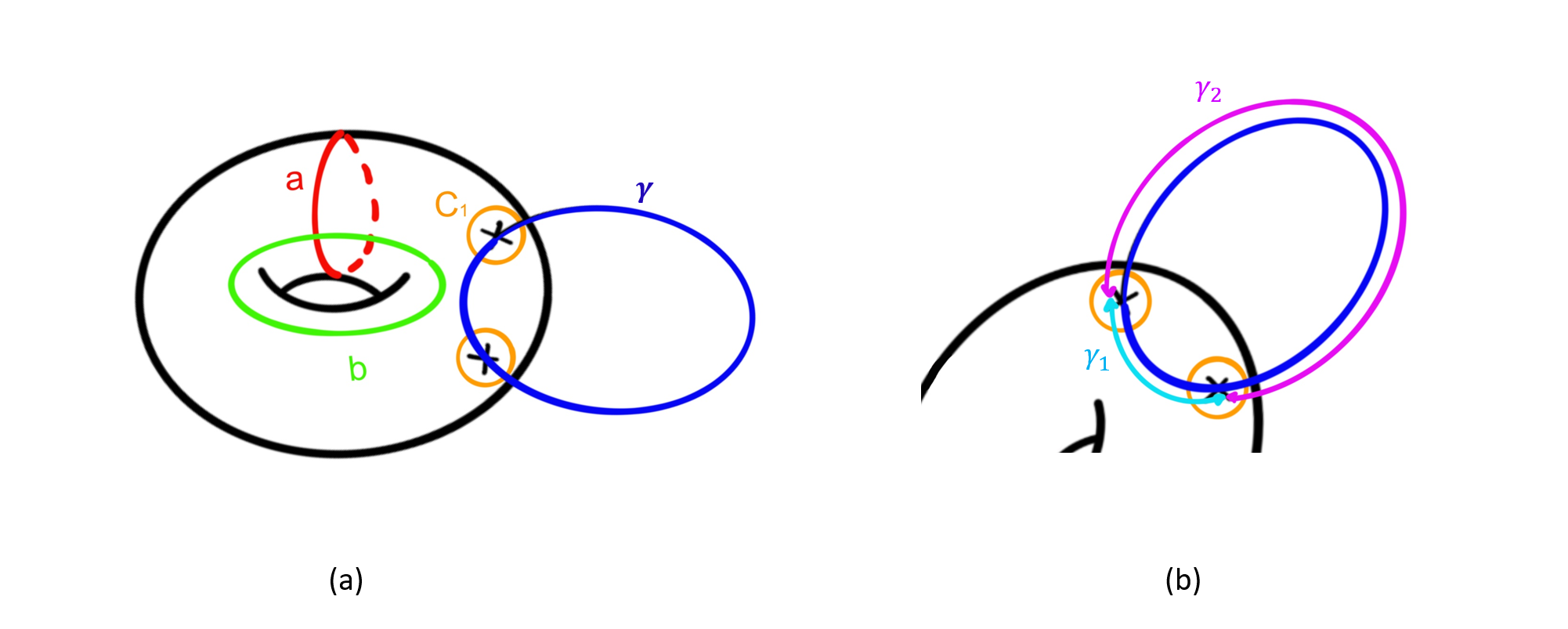}
    \caption{(a) Non-trivial cycles over $\tilde{\Sigma}_{1,2}$. (b) The curve $\gamma$ and its decomposition into the curves $\gamma_1$ and $\gamma_2$.}
    \label{fig:curves}
\end{figure*}

Defining,
\begin{align}
f\equiv \bigg(\frac{P_K}{\sqrt{W}}dX^K+\frac{P_H}{\sqrt{W}}dX^H+\frac{P_m}{\sqrt{W}} dX^m + P^+_0\bar{\Psi}\Gamma^-d\Psi\bigg),    
\end{align}
and as was shown in \cite{mpgm17}, the Hamiltonian is subject to five constraints. These are: a Local APD constraint

    \begin{equation}
    df= 0. \label{lc}
    \end{equation}
and four global APD constraints
\begin{align}
&\zeta_1 = \int_a f = 0,   \quad \zeta_2 = \int_b f = 0, \quad
&\zeta_3 = \int_{C_1} f = 0 \quad  \zeta_4 =  \int_{\gamma_1} f =0,   
\end{align}
where the first two are associated with the homology basis of cycles defined over the regular torus (see figure (\ref{fig:curves}-(a))). The one associated with the curve $C_1$ it is intrinsically related with singularities of the base manifold. The constraint associated with $\gamma_1$ (see figure (\ref{fig:curves}-(b))) is related with the homology curve along the deformed handle of $\Sigma_2$, which is still present after deforming $\Sigma_2$ into $\tilde{\Sigma}_{1,2}$.

In the following we will mention some of the fundamental features of the massive supermembrane Hamiltonian. From Eq. (\ref{MM}) it can be seen that it is very different from a standard compactification of the M2-brane on a $S^1$. Firstly, it contains a mass term associated with the non-trivial topology of the $LCD$ on the target space given by

\bea \label{fc}
\lim_{\epsilon\to 0} \int_{\Sigma^{'}} dK\wedge d\hat{H} \frac{\alpha \ m^2}{4}\{K,\hat{H}\}^2=2\pi\alpha m^2  k \hat{T}
\eea
This term can be interpreted as a as the uplift to ten non-compact dimensions of the central charge condition proposed in \cite{Restuccia}. In second place, it possesses non-vanishing mass terms associated with the dynamics fields $X^m$,$A^K$ and $A^H$, these are

\begin{eqnarray*}
        (\partial_KX^m)^2+(\ \partial_{\hat{H}} X^m)^2 &\not=0& , \\ (\partial_KA^K)^2+( \partial_{\hat{H}} A^K)^2 &\not=0& , \\ (\partial_KA^H)^2+(\partial_{\hat{H}} A^H)^2 &\not=0&.  
\end{eqnarray*}
Thus, the fermionic potential, been linear in the bosonic coordinates, is dominated by the bosonic potential due to these non-vanishing quadratic contribution to the Hamiltonian. This fact, together with the structure of the rest of the potential,  ensure that the Hamiltonian satisfy the  discreteness sufficient condition found in \cite{mpgm12}, as formerly shown in \cite{mpgm14}.

On the other hand, the supersymmetric analysis of this supermembrane formulation \cite{mpgm17}, shows that due to presence of punctures in the base manifold only half of the supersymmetry is preserved.

\section{Connections with Roman's massive supergravity}

In this section, we will discuss how the massive supermembrane can be related to Roman's massive supergravity. It is important to mention that our goal in this section is not to present a formal proof but instead to show a series of arguments that suggest a relation between the results presented here and Roman's supergravity as its low energy limit.

As discussed in \cite{mpgm14}, the singularities due to the punctures lead to $\delta^2$ singularities in the target space curvature. In detail, the metric (\ref{mono0}) near the punctures can be written as 

\begin{eqnarray}
    ds^2 \rightarrow l^2 \frac{|du_r|^2}{|u_r|^2}, \quad u_r = c_r|z-Z_r|\left(\frac{z-Z_r}{\bar{z}-\bar{Z}_r}\right)^{i\alpha/2l},
\end{eqnarray}
where $c_r$ are a constants with length units. Thus, near the punctures, the curvature satisfies the following condition 

\begin{eqnarray}
    R \rightarrow \hat{\delta}^2(|u_r|), 
\end{eqnarray}
where $\hat{\delta}^2(|u_r|)$ is defined as follows 

\begin{eqnarray} \label{deltadef1}
    && \int_0^{2\pi} d\theta_r \int_0^{U} d|u_r| \frac{2|u_r|}{l^2}\hat{\delta}^2(|u_r|) \phi(|u_r|) = 2\pi  \int_0^{U} d|u_r| \frac{2|u_r|}{l^2}\hat{\delta}^2(|u_r|) \phi(|u_r|) = - \phi(0). \nonumber \\
\end{eqnarray}
Here, we use $u_r = |u_r|e^{i\theta_r}$ and $\phi$ is a compact support function in the disk $|u_r|\leq U$. \newline 

On the other hand, near the zeros of $dF$, this is $P_a$, the line element (\ref{mono0}) can be written as

\begin{eqnarray}
    ds^2 & \rightarrow & l^2 \frac{|D(P_a)|^2|\tilde{u}_a|^2}{\tilde{c}_a^4}|d\tilde{u}_a|^2, \quad 2\frac{\tilde{u}_a^2}{\tilde{c}^2_a} = (z-P_a)^2+\beta_a (\bar{z}-\bar{P}_a)^2 + \frac{i\alpha}{2l}((z-P_a)^2-\beta_a (\bar{z}-\bar{P}_a)^2), \nonumber \\ &&
\end{eqnarray}
where $\tilde{c}_a$ are constants with length units and $\beta_a \equiv D(P_a)/\overline{D(P_a)}$. Hence, near the zeros, the curvature takes the form 

\begin{eqnarray}
    R \rightarrow \tilde{\delta}^2(|\tilde{u_a}|),
\end{eqnarray}
with 

\begin{eqnarray}
   && \int_0^{2\pi} d\tilde{\theta}_a \int_0^{\tilde{U}} d|\tilde{u}_a| \frac{2\tilde{c}^4}{l^2|D(P_a)|^2 |\tilde{u}_a|^3}\tilde{\delta}^2(|\tilde{u}_a|) \tilde{\phi}(|\tilde{u}_a|) \nonumber \\ && \quad \quad \quad = 2\pi \int_0^{\tilde{U}} d|\tilde{u}_a| \frac{2\tilde{c}^4}{l^2|D(P_a)|^2 |\tilde{u}_a|^3}\tilde{\delta}^2(|\tilde{u}_a|) \tilde{\phi}(|\tilde{u}_a|)  =  \tilde{\phi}(0).
\end{eqnarray}
As before, we use $\tilde{u}_a = |\tilde{u_a}|e^{i\tilde{\theta}_a}$ and $\tilde{\phi}$ is a compact support function in the disk $|\tilde{u}_a|\leq \tilde{U}$.

According to supergravity theory, $\delta$-type singularities imply the existence of p-branes sources \cite{Hull10}. In the case of the massive M2-brane the $\delta^2$ singularities suggest that the space-time source, at a supergravity level, must be of the type of the M9-brane presented in \cite{Bergshoeff7} (see also \cite{Sato}). This is the proposed source for the massive supergravity in eleven dimensions discussed in \cite{Bergshoeff6} as the uplift to eleven dimensions of Roman's massive supergravity. 

Let us consider, for example, a M9-brane simple solution of massive $D=11$ supergravity equation of motion given by \cite{Bergshoeff7,Sato}

\begin{equation}\label{M9sol}
    ds_{M9-brane}^2 = -H(y)^{-p/3} (dt^2-dx^2_8)+H(y)^{-10p/3-2}dy^2 +H(y)^{5p/3}dz^2,
\end{equation}
with

\begin{eqnarray}
    H(y) = d \pm \frac{\bar{m}}{p}|y|,
\end{eqnarray}
where $p$ and $d$ are constants. It is important to mention that this is a solution of the $D=11$ massive supergravity only when $p\not=0$. Moreover, in \cite{Bergshoeff6}, $z$ is a required isometry direction  and $\bar{m}$ is the Romans supergravity mass. This could be related to the M9-brane tension as following \cite{Sato},

\begin{eqnarray} \label{m-def}
    \frac{1}{16\pi G^{(11)}_N}\left(\int dz\right) \bar{m} = T_{M9},
\end{eqnarray}
where $T_{M9}$ is the M9-brane tension. It is important to mention that, by construction, the kinetic term of the M9-brane proposed in \cite{Bergshoeff7} describe a M9-brane wrapped along the isometry direction and therefore there is no integration in this direction (see Eq. (27) in \cite{Bergshoeff7}). Thus, the $T_{M9}$ could be directly identified as the D8-brane tension, $T_{D8}$, after a KK reduction along the isometry direction.

\begin{figure}
    \centering
    \includegraphics[scale=0.3]{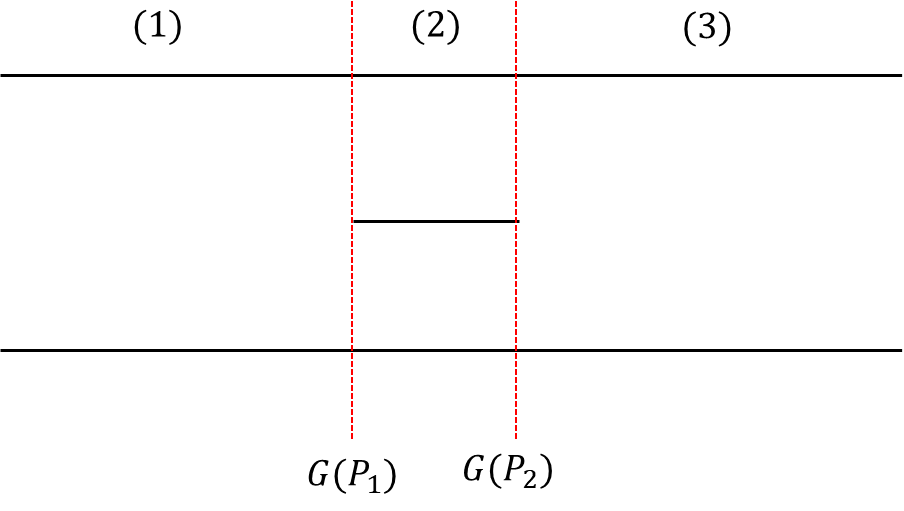}
    \caption{Region of the $LCD$ characterized by: (1) $G<G(P_1)$, (2) $G(P_1)<G<G(P_2)$ and (3) $G(P_2)<G$.}
    \label{fig:re}
\end{figure}

Now, it can be checked that near $y=0$ the curvature of the line element (\ref{M9sol}) can be written as 

\begin{eqnarray}
    R_{M_9} \rightarrow \bar{\delta}^2(y),
\end{eqnarray}

\begin{eqnarray}
 \int dz \int dy \frac{4}{3}H^{10p/3+1} \frac{\bar{m}^2}{16\pi G^{(11)}_N T_{M9}} \bar{\delta}^2(y) \bar{\phi}(y) = \int dy \frac{12}{9}H^{10p/3+1} \bar{m}\bar{\delta}^2(y) \bar{\phi}(y) = \bar{\psi}(0). \nonumber \\
\end{eqnarray}
Now, at this point, it is clear that the background metric used in this work for the massive supermembrane formulation does not coincide with the line element (\ref{M9sol}). This is not surprising since we are considering a target space that does not have the global isometry required for the massive supergravity formulation. However, we are only interested in comparing the singularities in order to show that the M9-branes can be the sources of our solution. Hence, we compare the singularities of the line element (\ref{M9sol}) with the ones get from the massive supermembrane background we are proposing (\ref{mono0}). In each case the curvature has a $\delta-type$ singularity, but with a different volumen element. Thus, we are looking the values for $d$ and $p$ in which the delta definition coincide. Let us begin with the singularities at the zeros of $dF$. Now, it is important to mention that in the cases of the zeros of $dF$, the comparison with the M9-brane solution only makes sense in the regions $G<G(P_1)$ and $G(P_2)<G$ (regions (1) and (3) in figure \ref{fig:re}). This is due to the fact that for $G(P_1)<G<G(P_2)$ (region (2)) we do not have the isometry direction required for the massive eleven-dimensional supergravity. This could indicate that, in this region, the Romans mass should be zero.

Thus, for the region (1) or (3), if $d=0$ and $p = -6/5$ it can be checked that the definition of $\bar{\delta}^2$ and $\tilde{\delta}^2$ coincide and we get the following relation between the massive supermembrane parameters and the Romans mass for each $P_a$

\begin{eqnarray}
    \bar{m}_a = \frac{1}{3}\left(\frac{5}{2}\right)^{3/2} \frac{l|D(P_a)|}{\sqrt{\pi}\tilde{c}_a^2}, \quad l=k\hat{T}.
\end{eqnarray}
In the same regions, (1) and (3), we can also identify $z \rightarrow H$ in equation (\ref{m-def}) to get the following relation

\begin{eqnarray}
    \frac{\alpha \bar{m}_a}{8G_N^{(11)}} = T^a_{M9}.
\end{eqnarray}
Thus in the limit to ten dimensions ($\alpha \rightarrow 0$) we find

\begin{eqnarray}
    \frac{\bar{m}_a}{8G_N^{(10)}} = T^a_{M8}.
\end{eqnarray}

In the case of the punctures, it can be checked that there are no values of $p$ and $d$ in which the definition of $\bar{\delta}^2$ and $\hat{\delta}^2$ coincide. However, if we consider the case $|p|<<1$ and $d=0$, we can write

\begin{eqnarray}
\frac{4}{3}H^{10p/3+1} \bar{m} = \frac{4\bar{m}^2|y|}{2p}\left(1+\frac{10}{3}p\ln\left(\frac{\bar{m}y}{p} \right) + \cdots \right).
\end{eqnarray}
Comparing this expression with  (\ref{deltadef1}) it is clear there is a match but only at leading order in $p$ when $|p|<<1$. In other words the metric proposed in this work for the massive supermembrane formulation is only describing part of the information contained in the metric \ref{M9sol}. Thus, we could write the following relation

\begin{eqnarray}
    \frac{4\bar{m}^2}{2p} = \frac{2}{l^2} + (\mbox{corrections in p}).
\end{eqnarray}
Hence, in order to find a complete match in these points, there are at least two possible ways in which one can proceed. The first one is to find another M9-brane solution considering an ansatz different than the one used in \cite{Bergshoeff7,Sato} to obtain (\ref{M9sol}). A way to do it, would be to consider different warp factors in the line element that could also have a dependence on $y,z$. A second way could be to address this problem by formulating the massive supermembrane directly in the background generated by multiple M9-branes using the line element (\ref{M9sol}) as shown in \cite{Bergshoeff7}. Now, this would lead to the highly non trivial problem of finding and analysing the complete form of the supersymmetric part of the supermembrane Hamiltonian \cite{deWit}. We will consider these points elsewhere.

Consequently, we found that the massive M2-brane \cite{mpgm14,mpgm17} is consistent at effective level with a 11D supergravity coupled to M9-branes. The source associated to Romans supergravity are D8-branes and in the 11D massive supergravity they correspond to the M9-branes.

\section{Massive type IIA superstring}
In this section, we analyze the double dimensional reduction of the massive supermembrane. We obtain a modified $N=2$ type IIA string, we denote it as \textit{massive} string. The first step of this reduction requires to obtain the string configurations of the massive supermembrane, where all dynamics fields only depend in one direction of the base manifold. One possibility is to consider that all the fields depend on a linear combination of $K$ and $H$, similar to the ansatz considered in \cite{mpgm18}, however, the different geometrical nature of the variables $K$ and $H$ make unclear the interpretation of the new variable obtained by this combination. The other possibility is to consider that all the supermembrane dynamics fields are of the form 

\begin{eqnarray} \label{az01}
X^m = f^m(H)\hat{X}^m(t,K) + U^m(H), \quad A^K = f^K(H)\hat{A}^K(t,K)+ U^K(H),
\end{eqnarray}

\begin{eqnarray} \label{az02}
A^H =  f^H(H)\hat{A}^H(t,K) + U^H(H), \quad \Psi = \lambda^H(H)\hat{\Psi}(t,K) + \Delta(H).
\end{eqnarray}
or 

\begin{eqnarray} \label{az001}
X^m = f^m(K)\hat{X}^m(t,H) + U^m(K), \quad A^K = f^K(K)\hat{A}^K(t,H)+ U^K(K),
\end{eqnarray}

\begin{eqnarray} \label{az002}
A^H =  f^H(K)\hat{A}^H(t,H) + U^H(K), \quad \Psi = f^{\Psi}(K)\hat{\Psi}(t,H) + U^{\Psi}(K).
\end{eqnarray}
Now, following the same argument presented in \cite{mpgm14}, it can be see that the previous expressions are not supersymmetric configurations of the massive supermembrane unless $f^m = f^K = f^H = f^{\Psi} = 1$ and all the $U's$ functions are proportional to $K$ (for (\ref{az01})-(\ref{az02})) or $H$ (for (\ref{az001})-(\ref{az002})). If all the $U's$ functions are proportional to $K$ or $H$ the 1-forms $dX^m$, $dA^K$,$dA^H$ and $d\Psi$ will be harmonic, which is a contradiction with the massive supermembrane formulation. Thus, all $U's$ functions should be constant which, without loss of generality, we can set to zero. 

This lead us only with two possibilities: the first one is that all the supermembrane fields are functions of $(t,K)$ and the second one when the fields are functions of $(t,H)$. Since we search for a closed \textit{massive} string by taking a limit in which we eliminate the only compact target-space direction, $H$, the only possibility is to assume that, all the fields are functions of $(t,K)$  
\begin{eqnarray} \label{az1}
X^m = \hat{X}^m(t,K), \quad A^K = \hat{A}^K(t,K),
\end{eqnarray}

\begin{eqnarray} \label{az2}
A^H =  \hat{A}^H(t,K), \quad \Psi = \hat{\Psi}(t,K).
\end{eqnarray}

Introducing these expressions in (\ref{gc}) we get the following result 

\begin{equation}\label{sgc}
   \partial_K \hat{A}^{H}=0, \quad \Rightarrow \quad \hat{A}^{H} = \hat{A}^{H}(t).
\end{equation}

Under this conditions the local supermembrane constraint (\ref{lc}) is trivially satisfied and from the global constraint $\zeta_3$

\begin{eqnarray}
\int_{C_1} \partial_t\hat{A}^H dH=0.
\end{eqnarray}
which implies that $\hat{A}^H$ is a constant,i.e. 

\begin{eqnarray}
\hat{A}^{H} = constant.
\end{eqnarray}
The global constraints $\zeta_1$ and $\zeta_2$ are trivially satisfied. At this point it is important to mention that, if we had assumed that all the fields were functions of $(t,H)$, the previous procedure would result in eliminating the dynamic field $A^K$, and therefore we would end with a geometrical object more similar to a string with one compact dimension. In the case we consider, in distinction, the 10D space is noncompact.

Continuing with our original assumption, we can write the string configurations Hamiltonian from the massive M2-brane one as
\begin{align}
H_{s} &= \frac{(4\pi l \alpha T_{M2} m)^2}{2P_0^+}+ \frac{2\pi}{P_0^+}\lim_{\epsilon \rightarrow 0}\int_{\Sigma'}d\hat{K}\wedge d\hat{H} [P_m^2 + P_K^2 + m^2 T^2_{M2}\alpha^2((\partial_{\hat{K}}\hat{X}^m)^2 + \alpha(\partial_{\hat{K}}\hat{A}^K)^2)\nonumber \\ &-2mT_{M2}P_0^+\bar{\Psi} \Gamma^-\Gamma_H\partial_{\hat{K}}\Psi].
\end{align}
Following the same procedure presented in \cite{mpgm14} (and more recently \cite{mpgm16}) the later expression can be written as 

\begin{align}
H_{s} &= \frac{( 4\pi l \alpha T_{M2} m)^2}{2P_0^+} + \frac{(2\pi)^2}{2P_0^+}\int_{-1}^{1}d\hat{K}[P_m^2 + P_K^2 + m^2T^2_{M2}\alpha^2((\partial_{\hat{K}}\hat{X}^m)^2 + (\partial_{\hat{K}}\hat{A}^K)^2) \nonumber \\ &-\frac{mT_{M2}\alpha P_0^+}{2\pi}\bar{\Psi}\Gamma^-\Gamma_H\partial_{\hat{K}}\Psi].
\end{align}
Now, setting 

\begin{eqnarray}
\theta = \frac{1}{2}(1+\hat{K})\pi
\end{eqnarray}
we get

\begin{align} \label{hal}
H_s &= \frac{ (4\pi l m \alpha T_{M2})^2}{2P_0^+}  + \frac{4\pi}{P_0^+} \int_0^\pi d\theta \bigg([P_{\hat{m}}^2 + \frac{(\pi mT_{M2}\alpha )^2}{4}(\partial_{\theta}\hat{X}^{\hat{m}})^2 - \frac{mT_{M2}\alpha P_0^+}{4}\bar{\Psi}\Gamma^-\Gamma_H\partial_{\theta}\Psi]\bigg).
\end{align}
where $\hat{X}^{\hat{m}}=(\hat{X}^m,\hat{A}^K)$. This Hamiltonian is restricted to the constraint $\zeta_4 = 0$, which can be rewritten as

\begin{eqnarray} \label{lmc}
\int_0^\pi \bigg(P_{\hat{m}}d\hat{X}^{\hat{m}}+P_0^+\bar{\Psi}\Gamma^-d\Psi\bigg) = 0.
\end{eqnarray}

Notice that, due to (\ref{mbc1}) and (\ref{mbc2}), this is a Hamiltonian for a closed string, and the constraint will be the corresponding level matching constraint. However, in order to fully establish this statement, we must eliminate the non-dynamic string represented by the $\gamma_2$ curve. Thus, since the string attached to $\Sigma_{1,2}$ does not contain any dynamical degrees of freedom, we will take the limit where the length of the curve $\gamma_2$ tends to zero. The string-like spike imposes the same boundary conditions on the two punctures, and it defines a closed string.

Now, let us analyse in detail the fermionic contribution of (\ref{hal}) in order to characterize that the string truly corresponds to the type IIA sector. For this, we use the following representation for the gamma matrix \footnote{Notice that the election for $\Gamma^K$ and $\Gamma^H$ is quite arbitrary. However, as is expected, the results will not depends on this.} 

\begin{eqnarray*}
    \Gamma^+ &=& \sqrt{2}\left(\begin{array}{cc}
        0 & 0 \\
        \mathbb{I}_{16} & 0 
    \end{array}\right), \quad \Gamma^- = \sqrt{2}\left(\begin{array}{cc}
        0 & \mathbb{I}_{16} \\
        0 & 0 
    \end{array}\right), \\  \Gamma^m &=& \left(\begin{array}{cc}
         \gamma^m  & 0 \\
          0 & -\gamma^m 
    \end{array}\right), \quad \Gamma^K = \left(\begin{array}{cc}
         \gamma^8  & 0 \\
          0 & -\gamma^8  
    \end{array}\right),  
     \quad \Gamma^H = \left(\begin{array}{cc}
         \gamma^9  & 0 \\
          0 & -\gamma^9  
    \end{array}\right). \label{GMR}
    \end{eqnarray*}
where $\gamma^{\hat{m}} = (\gamma^m,\gamma^8,\gamma^9)$ are the Euclidean $SO(9)$ gamma matrices. Now, we can solve the condition $\Gamma^+\Psi =0$ as 

\begin{eqnarray}
\Psi = k \left(\begin{array}{c}
     0  \\
     S 
\end{array}\right),
\end{eqnarray}
where $S$ is $16$ components Majorana spinor and $k$ is a constant.  The fermionic term in the Hamiltonian can be rewritten as

\begin{eqnarray}
\bar{\Psi}\Gamma^-\Gamma_H\partial_{\theta}\Psi = -\sqrt{2}k^2S^\dagger\gamma^9\partial_\theta S.
\end{eqnarray}
Now, without loss of generality, we can assume that we are in a representation were $\gamma^9$ is the chirality matrix in ten dimensions. Thus, we can decompose the spinors as $S = S^1 + S^2$, with $\gamma^9S^r = (-1)^{r+1}S^r$ and therefore 

\begin{eqnarray}
\bar{\Psi}\Gamma^-\Gamma_H\partial_{\theta} \Psi = \sqrt{2}k^2 [(S^2)^\dagger\partial_\theta S^2 - (S^1)^\dagger\partial_\theta S^1].
\end{eqnarray}
Notice that this term can be written in the same form as \cite{Green4}, that is 

\begin{eqnarray}
[(S^2)^\dagger\partial_\theta S^2 - (S^1)^\dagger\partial_\theta S^1] = \chi^\dagger \rho^0 \rho^1 \partial_\theta \chi,
\end{eqnarray}
where $\rho^0, \rho^1$ are the Dirac matrices in two dimensions 

\begin{eqnarray}
\rho^0 = \left(\begin{array}{cc}
    0 & -i \\
    i & 0
\end{array} \right), \quad \rho^1 = \left(\begin{array}{cc}
    0 & i \\
    i & 0
\end{array}\right),
\end{eqnarray}
and 
\begin{eqnarray}
\chi = \left( \begin{array}{c}
     S^2  \\
     S^1 
\end{array}\right).
\end{eqnarray}
Thus, the Hamiltonian can be written as \footnote{See that that $\chi^\dagger \rho^0$ is the definition of the Dirac conjugate on the worldsheet.}

\begin{eqnarray}
H_s &=& 4ml^2 T_{M2}\alpha+  \left(\frac{1}{4\pi(2\pi^2mT_{M2}\alpha)}\right) \int_0^\pi d\theta \bigg([\hat{P}_{\hat{m}}^2 + \left(2\pi^2mT_{M2}\alpha \right)^2(\partial_{\theta}\hat{X}^{\hat{m}})^2 \nonumber \\ &-& \frac{i}{\pi}\left(2\pi^2mT_{M2}\alpha \right) \chi^\dagger \rho^0 \rho^1 \partial_\theta \chi]\bigg),
\end{eqnarray}
where we set

\begin{eqnarray}
k^2 = \frac{i}{4\sqrt{2}m\pi^3 T_{M2}\alpha}, \quad P_0^+ =  2\pi^2 m T_{M2} \alpha, \quad \hat{P}_{\hat{m}}=4\pi P_m.
\end{eqnarray}

Following \cite{Duff} (see also \cite{Nicolai2}), we will define the string tension as 

\begin{eqnarray}
T_s = \alpha T_{M_2}.
\end{eqnarray}
Thus, to complete the double dimensional reduction we must take the limit $\alpha \rightarrow 0$ in such a way that $\alpha T_{M2}$ remains constant. Notice that, in our expressions, the string tension gets modified to 

\begin{eqnarray}
\tilde{T}_s = 2\pi^2m  T_{s}.
\end{eqnarray}
Although in the double dimensional reduction, the dependence on $H$ is lost, its behavior
around the punctures in terms of a characteristic integer $m\ne 0$, 
appears in the definitions of massive string tension, $\widetilde{T}_s$.
Then, the final Hamiltonian obtained by the double dimensional reduction of the massive supermembrane can be written as

\begin{eqnarray}
H_s &=& \frac{2l^2 \tilde{T}_s}{\pi^2} + \frac{1}{4\pi \tilde{T}_s} \int_0^\pi d\theta \bigg([\hat{P}_{\hat{m}}^2 + \tilde{T}_s^2(\partial_{\theta}\hat{X}^{\hat{m}})^2 - \frac{i}{\pi}\tilde{T}_s \chi^\dagger \rho^0 \rho^1 \partial_\theta \chi]\bigg),
\end{eqnarray}
and subject to 

\begin{eqnarray} \label{lmc}
\int_0^\pi \bigg( \hat{P}_{\hat{m}}\partial_\theta \hat{X}^{\hat{m}}+\frac{i}{2\pi}\chi^\dagger \partial_\theta\chi\bigg)d\theta = 0,
\end{eqnarray}

Now we can expand the maps as is usual in string theory, that is,

\begin{eqnarray}
    X^{\hat{m}} & = & x_0^{\hat{m}} + \frac{P_0^{\hat{m}}}{2\pi \tilde{T}_s}t + \frac{i}{2}\sqrt{\frac{1}{\pi \tilde{T}_s}}\sum_{n\not=0}\bigg[\frac{\alpha_n^{\hat{m}}}{n}e^{-2in(t-\theta)}+\frac{\Tilde{\alpha}_n^{\hat{m}}}{n}e^{-2in(t+\theta)}\bigg], \\
    S^1 &=& \sum_{n\not=0} S_n e^{-2in(t-\theta)}, \\
    S^2 &=& \sum_{n\not=0} \tilde{S}_n e^{-2in(t+\theta)},
\end{eqnarray}
where $x_0^{\hat{m}}$ and $P_0^{\hat{m}}$ are constants representing the mass center position and momentum. Using these expressions we can write the Hamiltonian as following 

\begin{eqnarray}
    H_s = \frac{2l^2\tilde{T}_s}{\pi^2} + \frac{1}{4\pi \tilde{T}_s}\sum_{\hat{m}}(P_0^{\hat{m}})^2 + 2(N + \bar{N}),
\end{eqnarray}
where $N$ and $\bar{N}$ are the number operators defined as

\begin{eqnarray}
    N = N_B + N_F, \quad  \bar{N} = \bar{N}_B + \bar{N}_F,
\end{eqnarray}
with

\begin{eqnarray}
    N_B = \sum_{\hat{m}}\sum_{n\not=0} \alpha_{-n}^{\hat{m}}\alpha_n^{\hat{m}}, \quad \bar{N}_B = \sum_{\hat{m}}\sum_{n\not=0} \tilde{\alpha}_{-n}^{\hat{m}}\tilde{\alpha}_n^{\hat{m}},
\end{eqnarray}

\begin{eqnarray}
    N_F =  \sum_{\hat{m}}\sum_{n\not=0} n S_{-n}^{\hat{m}}S_n^{\hat{m}}, \quad \bar{N}_F = \sum_{\hat{m}}\sum_{n\not=0}n \tilde{S}_{-n}^{\hat{m}}\tilde{S}_n^{\hat{m}}.
\end{eqnarray}
Thus, the mass operator can be written as 

\begin{eqnarray}
    M^2 \equiv 4\pi \tilde{T}_s H_s - \sum_{\hat{m}}(P_0^{\hat{m}})^2 = \frac{2l^2\tilde{T}^2_s}{\pi} + 8\pi \tilde{T}_s(N + \bar{N}).
\end{eqnarray}
Finally, the level matching constraint leads to 

\begin{eqnarray}
    N-\bar{N} = 0.
\end{eqnarray}
We have obtained the mass operator of a type IIA massive superstring in 10D, satisfying the standard level matching constraint for a string in ten non-compact dimensions. However, it exhibits features that make it different from the usual type IIA superstring mass operator. It possesses a non-vanishing constant term of topological origin in M-theory given in terms of a modified \textit{massive} string tension together with a dependence of the parameter $l$ on the Romans mass term. Although the massive string will be slightly modified when formulated on a more general background, the distinctive signature of Romans supergravity is already captured in the present description. It appears as a shift in the Hamiltonian description, fixing a constant scale in energy associated with the induced effect on the string of the non-trivial topology of space-time. At an effective level, it introduces a cosmological constant parametrized by the Romans mass term.
\section{Conclusions}

A striking goal in the literature has been the search of the Type IIA string worldsheet consistent with Romans supergravity. 

In this paper we obtain a worldsheet description of a massive type IIA superstring in ten noncompact dimensions space-time. Despite the simplicity of the background considered in this work, it still contains the mass terms characteristic of Romans supergravity. We perform a process of double dimensional reduction of the massive supermembrane obtained in \cite{mpgm14,mpgm17}. In the reduction process the dynamical maps can only be functions of $(t,K)$ or $(t,H)$. However, the only possibility to obtain a closed string Hamiltonian with ten non-compact directions in the target space is when all the fields are functions of $(t,K)$. After the dimensional reduction, we end up with a type IIA $N=2$ closed string Hamiltonian that contains a non-vanishing topological term and a modified string tension. The topological contribution is inherited from the topological term of the massive supermembrane and keeps track of the non-trivial structure of the Riemann surface. This inherited term also contains a parameter $T$ related to the Romans mass term. In addition it represent a shift in the spectrum of the string.

We obtain several connections between our formulation and a consistent uplift to M-theory of 10D Romans supergravity:\newline
In first place, the massive M2-brane \cite{mpgm13,mpgm14,mpgm17} was constructed as an explicit realization of Hull's conjecture in M-theory. When it is dimensionally reduced to $M_9\times T^2$ one obtains a M2-brane on a torus bundle with parabolic monodromy whose low energy description corresponds to type II parabolic gauged supergravity in 9D.  This is relevant because Romans supergravity when KK reduced  leads to a type II parabolic gauged supergravity in 9D. We also show a connection between the massive supermembrane and Roman's supergravity. It appears when we analyze the type of singularities of the twice punctured torus at a supergravity level. Indeed, it is possible to verify that the metric (\ref{mono0}) leads to $\delta^2$ singularities in the curvature scalar, which implies the presence of singularities at the supergravity level. The order of the singularities indicate that the source associated with these points should correspond to a wrapped M9-branes as the ones considered in \cite{Bergshoeff7}. Furthermore in \cite{mpgm17} we analyzed in detail the associated superalgebra of the massive M2-brane and we obtained that one of the surface terms realizes, on top of the M2-brane central charge, a central charge of the type $Z_{+M}$ whose dual has been associated with the presence of a 9-brane. Hence, the M2-brane background considered here is consistent with the presence at supergravity level of M9-branes, objects characteristic of the uplift to 11D of Romans supergravity.

Furthermore, in order to establish more evidences between the connection of the massive supermembrane formulation and the massive supergravity, we relate our results with those of  \cite{Bergshoeff6,Bergshoeff7}, at a supergravity level. We have considered the M9-brane solution to massive $D=11$ supergravity with the background metric of the massive supermembrane. We find that this M9-brane solution can reproduce the singularities of the massive M2-brane background in the zeros of $dF$. This allows us to identify the mass term of the cosmological constant at supergravity level (related to the presence of the M9-branes) with the parameter $T$ of the topological term present in the massive supermembrane as well as in the massive superstring formulations. In the punctures, we could not associate them to the particular M9-brane solutions considered in \cite{Bergshoeff7}, in order to reproduce the same type of singularity. However, we will like to stress that this does not imply that these singularities cannot be described in terms of M9-branes, only that this particular solution cannot. In order to establish a closer relationship between the massive M2-brane and Roman's supergravity, we need to consider the formulation of the M2-brane on a more general background.

In any case, massive supergravity in 11D requires the presence of M9-branes,- which in 10D are associated with D8 domain wall solutions-, and we have shown that their associated central charges require to couple to a supermembrane formulated on a punctured manifold (i.e. a massive supermembrane) like the one we have presented here, and in \cite{mpgm14,mpgm17}. 

Finally, we would like to comment that our construction can be extended to $N$ punctures with weight $\alpha_i$, such that $\sum_i \alpha_i =0$, and higher genus, g, surfaces. In this case, the Mandelstam map is also well defined, and we end up with a general Light Cone Diagram in the target space, $M_9\times LCD$. It coincides with the g-loop and N strings interaction diagram. In the formulation of the supermembrane we are considering, the moduli of this surface are fixed parameters. After the double dimensional reduction some of the moduli of the $LCD$ are captured  by the topological parameter and the tension of the \textit{massive} string. A possible generalization of our construction would be to consider instead of one $LCD$, the complete moduli space of punctured Riemann surfaces.

In conclusion, we have found a type IIA \textit{massive} closed string that represents a first step to the full-fledge massive type IIA string formulation of Roman's supergravity. In order to find a complete agreement, it may be necessary to explore more complicated backgrounds than the one used in this work (and in \cite{mpgm14,mpgm17}).

\section{Acknowledgements}

P.L. has been supported by the projects MINEDUC-UA ANT1956, ANT2156 and ANT2255 of the U. de Antofagasta. P.L is grateful with ANID/ POSTDOCTORADO BECAS CHILE/ 2022 - 74220031. MPGM is partially supported by the PID2021-125700NB-C21 MCI Spanish Grant. MPGM thanks to the Physics Department at Antofagasta U. for kind hospitality during the research stay, where part of this work was done. P. L was supported in part by a grant from the Gluskin She /Onex Freeman Dyson Chair in Theoretical Physics and by Perimeter Institute. Research at Perimeter Institute is supported in part by the Government of Canada through the Department of Innovation, Science and Industry Canada and by the Province of Ontario through the Ministry of Colleges and Universities.

\bibliographystyle{unsrt}







\end{document}